\documentclass[10pt,twocolumn,aps,pra,showpacs,superscriptaddress,floatfix,nofootinbib]{revtex4-1}

\usepackage{graphicx,tikz}
\usepackage{amssymb}
\usepackage{amsmath}
\usepackage{color,xcolor}
\usepackage{psfrag}
\usepackage{epsfig}
\usepackage{bbm}
\usepackage{bm}
\usepackage{hyperref}
\usepackage[normalem]{ulem}
\usepackage{amsthm}
\usepackage{epstopdf}
\usepackage{xcolor}
\usepackage{times}
\usepackage{color,soul}
\usepackage{tcolorbox}
\usepackage{dirtytalk}
\usepackage{csquotes}

\newcommand{\Q}{\mathcal{Q}}
\newcommand{\AQ}{\tilde{\Q}}
\newcommand{\NS}{\mathcal{NS}}

\newcommand{\T}{\mathcal{H}}

\newcommand{\hT}{\mathfrak{H}}
\newcommand{\hL}{\mathfrak{L}}
\newcommand{\hQ}{\mathfrak{Q}}
\newcommand{\hAQ}{\tilde{\mathfrak{Q}}}
\newcommand{\hNS}{\mathfrak{N}}

\newcommand{\vecf}{\vec{f}}
\newcommand{\vecP}{\vec{P}}
\newcommand{\DKL}{D_\text{\tiny KL}}

\newcommand{\Dkl}{\DKL(\vecf||\T)}

\newcommand{\PKL}[1]{\vecP_\text{\tiny KL}^{#1,*}}
\newcommand{\Nest}{N_{\text{est}}}
\newcommand{\Ntot}{N_{\text{total}}}

\newcommand{\id}{\mathbb{I}}
\newcommand{\SCHSH}{\mathcal{S}_{\mbox{\tiny CHSH}}}

\newcommand{\vecPs}{\vecP(v,\epsilon,\{p_j\})}
\newcommand{\vecPv}{\vecP(v)}

\newcommand{\vecPExt}{\vecP^\text{\tiny Ext}_j}

\DeclareMathOperator{\tr}{tr}

\renewcommand{\L}{\mathcal{L}}

\begin{document}
\title{Bounding the plausibility of physical theories in a device-independent setting via hypothesis testing}

\author{Yeong-Cherng Liang}
\email{ycliang@mail.ncku.edu.tw}
\affiliation{Department of Physics and Center for Quantum Frontiers of Research \& Technology (QFort), National Cheng Kung University, Tainan 701, Taiwan}
\author{Yanbao Zhang}
\email{yanbaoz@gmail.com}
\affiliation{NTT Basic Research Laboratories and NTT Research Center for Theoretical Quantum Physics, NTT Corporation, 3-1 Morinosato-Wakamiya, Atsugi, Kanagawa 243-0198, Japan}

\date{ \today}

\begin{abstract}
The device-independent approach to physics is one where conclusions about physical systems (and hence of Nature) are drawn directly and solely from the observed correlations between measurement outcomes. This operational approach to physics arose as a byproduct of Bell's seminal work to distinguish, via a Bell test, quantum correlations from the set of correlations allowed by local-hidden-variable theories. In practice, since one can only perform a finite number of experimental trials, deciding whether an empirical observation is compatible with some class of physical theories will have to be carried out via the task of hypothesis testing. In this paper, we show that the prediction-based-ratio method---initially developed for performing a hypothesis test of local-hidden-variable theories---can equally well be applied to test many other classes of physical theories, such as those constrained only by the nonsignaling principle, and those that are constrained to produce any of the outer approximation to the quantum set of correlations  due to Navascu\'es-Pironio-Ac\'{\i}n.  We numerically simulate Bell tests using hypothetical nonlocal sources of correlations to illustrate the applicability of the method  in both the independent and identically distributed (i.i.d.) scenario and the non-i.i.d. scenario. As a further application, we demonstrate how this method allows us to unveil an apparent violation of the nonsignaling conditions in certain experimental data collected in a Bell test. This, in turn, highlights the importance of the randomization of measurement settings, as well as a consistency check of the nonsignaling conditions in a Bell test.
\end{abstract}
\pacs{}

\maketitle

%%%%%%%%%%%%%%%%%%%%%%%%%%%%%%%%%%%%%%%%%%%%%%%%%%%%%%%%%%%%%%%%%%%%%%%%%
%%%% Introduction

\section{Introduction}

In physics, the terminology ``device-independent'' apparently made its first appearance in Ref.~\cite{Acin:PRL:2006} where the authors drew a connection between the celebrated discovery by Bell~\cite{Bell64} and the vibrant field of quantum cryptography~\cite{Gisin:RMP:QCrypto}. As of today,  device-independent quantum information has become a well-established research area where Bell-inequality-violating correlations find  applications not only in the distribution of secret keys~\cite{Barrett05,Acin07,Vazirani14} (see also Ref.~\cite{Ekert91}), but also in the generation of random bits~\cite{Colbeck09,Pironio10,Colbeck11}, as well as in the assessment of uncharacterized devices  (see, e.g., Refs.~\cite{Mayers04,Brunner08,Reichardt13,Yang14,Liang15,Coladangelo17,Sekatski2018}). For a comprehensive review, see Refs.~\cite{Scarani12,Brunner14}.

A device-independent approach to physics, however, could be traced back, for example, to the work of Bell~\cite{Bell64}. There, he showed that {\em any} local-hidden-variable (LHV)  theory~\cite{Bell04} must be incompatible with certain quantum predictions. The proof is ``device-independent'' in the sense that one needs no further assumption about the nature of  the theory (including the detailed functioning of any devices that one may use to test the theory). Rather, the proof relies on a common ingredient of operational physical theories---correlations between measurement outcomes, i.e., the probability of getting particular  measurement outcomes conditioned on certain measurement choices being made---to manifest the incompatibility.

By now, this incompatibility has been verified in various loophole-free Bell tests, such as those reported in Refs.~\cite{Hensen15,Shalm15,Giustina15,rosenfeld_event-ready_2017,LiPRL2018}. Importantly, any real experiments must involve only a finite number of experimental trials. Statistical fluctuations must thus be carefully taken into account in order to draw any conclusion against a hypothetical theory, such as an LHV theory. For example, using the observed relative frequencies as a na\"ive estimator of the underlying correlations would generically (see, e.g., Refs.~\cite{Schwarz16,Lin2018}) lead to a violation of the nonsignaling conditions~\cite{Popescu1994,Barrett:PRA:2005}.  Since the assumption of nonsignaling is a prerequisite for any Bell tests, it is only natural that a Bell test of LHV theories must also be accompanied by the corresponding test of this assumption~\cite{Shalm15,Giustina15,rosenfeld_event-ready_2017,LiPRL2018,Liu:2018aa} (see also Refs.~\cite{Guillaume2017,Bednorz:PRA:2017,Kupczynski:2017}).

The effects of statistical fluctuations in a Bell test were (in fact, still are) often reported in terms of the number of standard deviations the estimated Bell violation exceeds the corresponding local bound (see, e.g., Refs.~\cite{Aspect:PRL:1982,Tittel:PRL:1998,Weihs:PRL:1998,Rowe01,Giustina:2013aa,Christensen:PRL:2013,Erven:2014aa,Lanyon14,Shen:PRL:2018}). However, there are several problems with such a statement (see Refs.~\cite{Brunner14, Zhang2011} for detailed discussions). Alternatively, as a common practice in hypothesis testing, one could also present the $p$-value according to a certain null hypothesis (e.g., the hypothesis that a LHV theory holds true). The corresponding $p$-value then describes the probability that the statistical model (associated with the null hypothesis) produces some quantity (e.g., the amount of Bell-inequality violation) at least as extreme as that observed.

A pioneering work in this regard is that due to Gill~\cite{Gill:2003} where he presented a $p$-value upper bound according to the hypothesis of a LHV theory based on the violation of the Clauser-Horne-Shimony-Holt  (CHSH)~\cite{Clauser69} Bell inequality. A few years later, a systematic method that works directly on the observed data (without relying on any predetermined Bell inequality)---by the name of the {prediction-based-ratio method}---was developed by one of the present authors and coworkers~\cite{Zhang2011} (see also Ref.~\cite{Zhang2013}). {This method} was designed for computing a $p$-value upper bound---based on the data collected in a Bell test---according to LHV theories. As we shall show in this work, essentially the same method can be applied for  the hypothesis testing of some other nonlocal physical theories, thus allowing us to bound the plausibility of physical theories beyond LHV theories. 

Indeed, since the pioneering work by Popescu and Rohrlich~\cite{Popescu1994}, there has been an ongoing effort~(see, e.g., Refs.~\cite{Cavalcanti:2010aa,Fritz:2013aa,Amaral:PRA:2014,Navascues:2015aa})  to find well-motivated physical~\cite{Navascues:2010aa,Rohrlich:2014aa} or information-theoretic~\cite{Dam:2013aa,Brassard:PRL:2006,Linden:PRL:2007,Pawowski:2009aa} principles to recover precisely the set of quantum correlations. Unfortunately, none of these has succeeded. Rather, they each define a set of correlations that outer approximates the quantum set~\cite{Goh2018}. In other words, they also contain correlations that are more nonlocal than that allowed by quantum theory. For example, the so-called  ``almost-quantum''~\cite{Navascues:2015aa}  set of correlations is one such superset of the quantum set, yet satisfying essentially all the proposed principle known to date. In the rest of this work, it suffices to think of this set as a fairly good outer approximation to the quantum set of correlations. 

In this work, we show that the {prediction-based-ratio} method can be applied to test any physical theory that is constrained to produce correlations that is amenable to a semidefinite programming~\cite{BoydBook} characterization.  In particular, it can be applied to test  any physical theory that is constrained to produce nonsignaling~\cite{Popescu1994}  correlations, or any theory that respects macroscopic locality~\cite{Navascues:2010aa} or which gives rise to the almost-quantum~\cite{Navascues:2015aa} set of correlations { etc.}

 %%%%%
\section{Methods}
\label{Sec:PBR}

\label{Sec:PBR}

\subsection{Preliminaries}

For a complete description of the {prediction-based-ratio} method and a comparison of its strength against the martingale-based method~\cite{Gill:2003}, we refer the reader to Ref.~\cite{Zhang2011}. Here, we merely recall the necessary ingredients of the prediction-based-ratio method and show how it can be used to achieve the purpose of bounding the plausibility of physical theories based on the data collected in a Bell test, with {\em minimal} assumptions.  Making this possibility evident and demonstrating how well it works in practice are the main contributions of the present work.

For simplicity, the following discussions are based on a Bell test that involves two parties (Alice and Bob) who are each allowed to perform one of two measurements randomly selected at each trial, each produces one of two possible outcomes. Generalization  to other Bell scenarios will be evident. To this end, let us denote the measurement choice (input) of Alice (Bob) by $x$ ($y$) and the corresponding measurement outcome (output) by $a$ ($b$), where $a,b,x,y\in\{0,1\}$. The extent to which the distant measurement outcomes are correlated is then succinctly summarized by the collection of joint conditional probability distributions $\vecP=\{P(a,b|x,y)\}{_{a,b,x,y}}$.

In an LHV  theory, the outcome probability distributions can be produced with the help of some  LHV  $\lambda$ (distributed according to $q_\lambda$) via the local response functions satisfying $0\le P^A_\lambda(a|x), P^B_\lambda(b|y)\le 1$ and $\sum_a P^A_\lambda(a|x)=\sum_b P^B_\lambda(b|y)=1$ such that~\cite{Bell64}:
\begin{equation}\label{Eq:Local}
	P(a,b|x,y) = \sum_\lambda q_\lambda P^A_\lambda(a|x) P^B_\lambda(b|y).
\end{equation}
Hereafter, we refer to any $\vecP$ that can be decomposed in the above manner as a (Bell-) local correlation and denote the set of such correlations as $\L$.

In contrast, if Alice and Bob conduct the experiment by performing local measurements on some shared quantum state $\rho$, quantum theory predicts setting-dependent outcome distributions for all $a,b,x,y$ of the form:
\begin{equation}\label{Eq:Q}
	P(a,b|x,y) = \tr(\rho\,M^A_{a|x}\otimes M^B_{b|y}),
\end{equation}
where $M^A_{a|x}$ and $M^B_{b|y}$ denote, respectively, the local positive-operator-value-measure element associated with the $a$-th outcome of  Alice's $x$-th measurement and the $b$-th outcome of Bob's $y$-th measurement. Accordingly, we refer to any $\vecP$ that can be written in the form of Equation~\eqref{Eq:Q} as a quantum correlation and the set of such correlations as $\Q$.

Importantly, both local and quantum correlations satisfy the nonsignaling conditions~\cite{Barrett:PRA:2005}:
\begin{equation}\label{Eq:NS}
\begin{split}
	P_A(a|x,y)=P_A(a|x,y'):=P_A(a|x)\quad\forall\,a,x,y,y',\\
	P_B(b|x,y)=P_B(b|x',y):=P_B(b|y)\quad\forall\,b,x,x',y,
\end{split}
\end{equation}
where $P_A(a|x,y):=\sum_b P(a,b|x,y)$ and $P_B(b|x,y):=\sum_a P(a,b|x,y)$ are marginal probability distributions of $P(a,b|x,y)$. Should (any of) these conditions be violated in a way that is independent of spatial separation, Alice and Bob would be able to communicate faster-than-light~\cite{Popescu1994} via the choice of measurement $x,y$. We shall denote the set of $\vecP$ satisfying Equation~\eqref{Eq:NS} as $\NS$. It is known that $\L$, $\Q$, and $\NS$ are convex sets and that they satisfy the strict inclusion relations $\L\subset\Q\subset\NS$ (see, e.g., Ref.~\cite{Brunner14} and references therein).

A few other convex sets of correlations are worth mentioning for the purpose of subsequent discussions. To this end, note that the problem of deciding if a given $\vecP$ is in $\Q$ is generally a difficult problem. However, the characterization of $\Q$ can, in principle, be achieved by solving a converging hierarchy of semidefinite programs~\cite{BoydBook}  due to Nacascu\'es, Pironio, and Ac\'in (NPA)~\cite{NPA,NPA2008} (see also Ref.~\cite{Doherty08,Moroder13}). The lowest level outer approximation of $\Q$ in this hierarchy, often denoted by $\Q_1\supset\Q$, happens to be exactly the set of correlations that is characterized by the physical principle of macroscopic locality~\cite{Navascues:2010aa}. A finer outer approximation of $\Q$ corresponding to the lowest-level hierarchy of Ref.~\cite{Moroder13}, which we denote by $\AQ$, is known in the literature as the almost-quantum set~\cite{Navascues:2015aa}, as it appears to satisfy all the physical principles that have been proposed to characterize $\Q$. In Section~\ref{Sec:Results}, we use $\AQ$ and $\NS$ as examples to illustrate how the {prediction-based-ratio} method can be adapted to test physical theories that are constrained to produce correlations from these sets. 

\subsection{Finite Statistics and the Prediction-Based-Ratio Method}

Coming back to an actual Bell test, let $\Ntot$ be the total number of experimental trials carried out during the course of the experiment. During each experimental trial, $x$ and $y$ are to be chosen randomly according to some fixed probability distribution $P_{xy}$. (This distribution may  be varied from one trial to another but for simplicity of discussion, we consider  in this work only the case where this is fixed once and for all before the experiment begins.) From the data collected in a Bell test, a na\"ive (but very commonly-adopted) way to estimate the correlation $\vecP$ between measurement outcomes is to compute the relative frequencies $\vecf$ that each combination of outcomes $(a,b)$ occurs given the choice of measurement $(x,y)$, i.e., 
\begin{equation}
	f(a,b|x,y) = \frac{N_{a,b,x,y}}{N_{x,y}},
\end{equation}
where $N_{a,b,x,y}$ is the total number of trials the events corresponding to $(a,b,x,y)$ are registered and $N_{x,y}=\sum_{a,b} N_{a,b,x,y}$ is the number of times the particular combination of measurement settings $(x,y)$ is chosen. By definition, $\Ntot=\sum_{x,y} N_{x,y}$. 

If the experimental trials are  independent and identically distributed (i.i.d.) corresponding to a {\em fixed} state $\rho$ with {\em fixed} measurement strategies $\{M^A_{a|x}\}_{a,x},\{M^B_{b|y}\}_{b,y}$, then in the asymptotic limit, $\lim_{\Ntot\to\infty} f(a,b|x,y) = P(a,b|x,y)$ where $\vecP$ here would satisfy Equation~\eqref{Eq:Q}. In this limit, the amount of statistical evidence in the data against a particular hypothesis $\mathfrak{H}$ can be quantified by the Kullback-Leibler (KL) divergence~\cite{Kullback1951} (also known as the relative entropy) from $\vecP$ to $\L$, see Refs.~\cite{vanDam:IEEE:2005,Acin:PRL:2005} for a detailed explanation with quantum experiments. We remark that the KL divergence is directly related with the Fisher information metric and so it measures the distinguishability of a distribution from its neighborhood. This provides a motivation for using the KL divergence as a measure of statistical evidence.

In the (original) {prediction-based-ratio} method of Ref.~\cite{Zhang2011} (see also Ref.~\cite{Zhang2010}), the hypothesis of interest is that the experimental data can be produced using an  LHV  theory, in other words, that the underlying correlation $\vecP\in\L$. For convenience, we shall refer to this hypothesis  as $\hL$. In this case, given $\vecf$ and $P_{xy}$, the relevant KL divergence from $\vecf$ to $\mathcal{L}$ reads as
\begin{equation}\label{Eq:KL}
	\DKL\left(\vecf|| \mathcal{L}\right):= \,\min_{\vecP\in\mathcal{L}}\sum_{a, b, x, y}  P_{xy} f(a, b | x, y) \log \left[ \frac{f (a, b | x, y)}{P (a, b | x, y)} \right]
\end{equation}
As the objective function in Equation~\eqref{Eq:KL} is strictly convex in $\vecP$ and the feasible set $\mathcal{L}$ is convex, the minimizer of the above optimization problem---which we shall denote by $\PKL{\L}$---is {\em unique} (see, e.g., Ref.~\cite{Lin2018}). It follows from the results presented in Ref.~\cite{Zhang2011} that this unique minimizer $\PKL{\L}$ can be used to construct a Bell inequality:
\begin{subequations}\label{Eq:OptBellIneq}
\begin{equation}
	\sum_{a,b,x,y} R(a,b,x,y)P_{xy}P(a,b|x,y)\stackrel{\mathcal{L}}{\le} 1,
\end{equation}
where the non-negative coefficients of the Bell inequality are defined via the ratios
\begin{equation}
	R(a,b,x,y):= \frac{f(a,b|x,y)}{\PKL{\L}(a,b|x,y)}.
\end{equation}
\end{subequations}
This Bell inequality is the key ingredient of the {prediction-based-ratio method} and is ideally suited for performing a hypothesis test of $\hL$. 

To understand {the method}, we introduce the random variables $X$ and $Y$ to denote the random inputs and the variables $A$ and $B$ to denote the random outputs of Alice and Bob at a trial. The ability to select measurement settings randomly, in particular, is an indispensable prerequisite of the {prediction-based-ratio} method, or more generally, a proper Bell test (see, e.g., Ref.~\cite{Bell04}). We further denote the possible values of inputs and outputs by the respective lower-case letters. Then we can think  of the ratio $R$ in Equation~\eqref{Eq:OptBellIneq} as a non-negative function of the inputs $X, Y$ and outputs $A, B$ at each experimental trial such that its expectation according to an arbitrary $\vecP \in \L$ with the fixed input distribution $P_{xy}$ satisfies 
\begin{equation}\label{def:pbr}
    \langle R(A, B, X, Y)\rangle \stackrel{\mathcal{L}}{\le} 1.
\end{equation}
Equation~\eqref{def:pbr} is an alternative way of expressing the Bell inequality of Equation~\eqref{Eq:OptBellIneq}. A real experiment necessarily involves only a finite number $\Ntot=(N_{\textrm{est}}+N_{\textrm{test}})$ of experimental trials in time order. Here, we have split the experimental data into two sets: the data from the first $N_{\textrm{est}}$ trials as the \emph{training data} and the data from the remaining $N_{\textrm{test}}$ trials as the \emph{hypothesis-testing data}.  In practice, we first construct the function $R$ using the training data and then perform a hypothesis test with the test data. Since the ratio $R$ is determined before 
the hypothesis test based on the prediction according to the training data, $R$ is called a {\emph{prediction-based ratio}}. 

Given a {prediction-based ratio} and a finite number $N_{\textrm{test}}$ of test data, we can quantify the evidence against the hypothesis $\hL$ by a $p$-value. For concreteness, suppose that the actual measurements chosen at the $i$-th test trial are $x_i$, $y_i$ and the corresponding measurement outcomes observed are $a_i$, $b_i$. Then the value of the {prediction-based ratio} at the $i$-th test trial is $R(a_i,b_i,x_i,y_i)$, abbreviated as $r_i$. We introduce a test static $T$ as the product of the possible values of the {prediction-based ratio} at all test trials, so the observed value of the test statistic is $t=\prod_{i=1}^{N_{\textrm{test}}} r_i$.
If we denote by $N'_{a,b,x,y}$ the total number of counts registered for the input-output combination $(a,b,x,y)$ in the test data, then $t$ can be expressed also as
\begin{equation}\label{Eq:T:simplified}
	t=\prod_{a,b,x,y} R(a,b,x,y)^{N'_{a,b,x,y}}. 
\end{equation}
According to Ref.~\cite{Zhang2011}, the $p$-value, which is defined as the maximum probability according to the hypothesis $\hL$ of obtaining a value of $T$ at least as high as $t$ actually observed in the experiment, is bounded by 
\begin{equation} \label{Eq:p-value_bound}
    p\leq \min \{1/t, 1\}. 
\end{equation}
The smaller the $p$-value, the stronger the evidence against the hypothesis $\hL$ is, in other words, the less plausible LHV theories are. It is worth noting that the $p$-value bound computed in this manner remains valid even if the experimental trials are not i.i.d., while when the experimental trials are i.i.d., the $p$-value bound is asymptotically optimal (or tight)~\cite{Zhang2011}.

\subsection{ Generalization for Hypothesis Testing Beyond LHV Theories}
\label{Sec:GeneralizedPBR}

The following two simple observations,  which allow one to apply the {prediction-based-ratio} method to test physical theories beyond those described by LHV, are where our novel contribution enters. Firstly, we make the observation that in the above arguments leading to the $p$-value bound of Equation~\eqref{Eq:p-value_bound}, the actual hypothesis $\hL$ only enters at Equation~\eqref{Eq:OptBellIneq} via the set of correlations $\L$ compatible with the hypothesis $\hL$. In particular, if we are to consider the hypothesis  $\hT$ that the data observed is produced by a physical theory H (e.g., a nonsignaling theory), then we merely have to replace $\L$ by the (convex) set of correlations $\T$ (e.g., $\NS$) associated with H in the optimization problem of Equation~\eqref{Eq:KL}. {The method} then allows us to bound the plausibility of the hypothesis $\hT$ via the $p$-value bound in Equation~\eqref{Eq:p-value_bound} with the possible values of the {prediction-based ratio} given by
\begin{equation}\label{Eq:R:T}
	R(a,b,x,y):= \frac{f(a,b|x,y)}{\PKL{\T}(a,b|x,y)},
\end{equation}
where $\PKL{\T}$ is the unique minimizer of the optimization problem:
\begin{equation}\label{Eq:KL:T}
	\DKL\left(\vecf|| {\T}\right):= \,\min_{\vecP\in\T}\sum_{a, b, x, y}  P_{xy} f(a, b | x, y) \log \left[ \frac{f (a, b | x, y)}{P (a, b | x, y)} \right].
\end{equation}

Although Equation~\eqref{Eq:T:simplified}, Equation~\eqref{Eq:p-value_bound} and Equation~\eqref{Eq:R:T} together provide us, in principle, a recipe to test  the plausibility of a general physical theory H, its implementation depends on the nature of the set of correlations associated with the hypothesis. Indeed, a crucial part of the procedure is to solve the optimization problem of Equation~\eqref{Eq:KL:T} for the convex set of correlations $\T$ compatible with H, which is generally far from trivial. If $\T$ is a convex polytope, such as $\L$ and $\NS$, or the set of correlations associated with the models considered in Refs.~\cite{Bancal:2012aa,Barnea2013}), it is known~\cite{Zhang2011} that Equation~\eqref{Eq:KL:T} can indeed be solved numerically. 

Our second observation is that for the convex sets of correlations that are amenable to a semidefinite programming characterization, such as those considered in Refs.~\cite{NPA,Moroder13,SLChen16,Chen2018}, Equation~\eqref{Eq:KL:T} is an instance of a conic program~\cite{BoydBook} that can be efficiently solved using  a freely available solver, such as PENLAB~\cite{PENLAB}. To see this, one first notes that, apart from the constant factor $P_{xy}$, the optimization of Equation~\eqref{Eq:KL:T}  is essentially the same as that considered in Ref.~\cite{Lin2018}.  A straightforward adaptation of the argument presented in Appendix D 2 of Ref.~\cite{Lin2018} would  then allow us to complete the aforementioned observation. The data observed in a Bell test can thus be used to test not only $\hL$, but also $\hNS$ and even the hypothesis $\hQ$ that the observation is compatible with Born's rule, cf. Eq.~\eqref{Eq:Q}, via outer approximations of $\Q$ (such as $\Q_1$ and $\tilde{\Q}$). 

A remark is now in order. In order to avoid so-called $p$-value hacking, it is essential that the test data used in the computation of the test statistic $T$ is not used to determine $\vecf$, and hence the values of the {prediction-based ratio} $R$ in Equation~\eqref{Eq:R:T}. In this work, for simplicity we use the first $N_{\textrm{est}}$ trials of an experiment as the training data for estimating $\vecf$ and further constructing a {prediction-based ratio} $R$ that is applied for all test trials. In principle, we can use different training data for different test trials. For example, we can define the training data for a test trial as the data from all trials performed before this test trial, and then we can adapt the construction of the {prediction-based ratio} for each individual test trial. We refer to Ref.~\cite{Zhang2011} for more details on the adaptability of the {prediction-based ratio}.

%%%%% Results
\section{Results}
\label{Sec:Results}

To illustrate how  well the {prediction-based-ratio method}  works in identifying data that are {\em not} even explicable by some nonlocal physical theories, such as quantum theory, we now consider a few examples of applications of the  method. As above, we restrict our attention to a bipartite Bell test, where each party performs two binary-outcome measurements randomly selected at each trial. Throughout this section, we assume that the input distribution is uniform, specifically $P_{xy}=\frac{1}{4}$ for all combinations of $x,y\in\{0,1\}$. In Section~\ref{Sec:iid} and Section~\ref{Sec:Niid} we study the behaviour of numerically simulated  Bell tests based on hypothetical sources of correlations described in Section~\ref{Sec:Model}, while in Section~\ref{Sec:RealData}, we analyze the real experimental data reported in Ref.~\cite{Christensen15}.

\subsection{Modeling a Bell Test}
\label{Sec:Model}

For our numerical simulations, we consider a $\vecP$ that resembles a nonlocal source targeted at in various actual Bell experiments~\cite{Tittel:PRL:1998,Weihs:PRL:1998,Rowe01,Christensen15,Poh15}:
\begin{equation}\label{Eq:vecPv}
	\vecP(v):=v\vecP_\text{\tiny PR} +(1-v)\vecP_{\mathbb{I}},
\end{equation}
where $v\in[0,1]$, $\vecP_\text{\tiny PR}$ is the Popescu-Rohrlich (PR) correlation~\cite{Popescu1994} $P_\text{\tiny PR}(a,b|x,y) = \tfrac{1}{2}\delta_{a\oplus b,xy}$ with $a,b,x,y\in\{0,1\},$ and $P_{\id}(a,b|x,y)=\tfrac{1}{4}$ for all $a,b,x,y$ is the white-noise distribution. In Equation~\eqref{Eq:vecPv},  the real parameter $v$  can be seen as the weight associated with $\vecP_\text{\tiny PR}$ in the convex mixture. Importantly, the nonlocal source represented by such a mixture can (in principle) be produced by performing appropriate local measurements on a maximally entangled two-qubit state if and only if $v\le v_c:=\frac{1}{\sqrt{2}}\approx 0.71$ (see, e.g., Refs.~\cite{Lin2018,Goh2018}). In particular, when $v=v_c$---corresponding to an ideal nonlocal source---the mixture gives rise to the maximal quantum violation of the CHSH~\cite{Clauser69} Bell inequality. 

To mimic an experimental scenario with noise (something unavoidable in practice), we shall introduce a slight perturbation to the ideal source $\vecP(v)$ of Equation~\eqref{Eq:vecPv}. Specifically, we require the measurement outcomes observed at each trial in the simulated Bell test to be governed by the nonlocal source $(1-\epsilon)\vecP(v) + \epsilon\vecP_\text{\footnotesize noise}$, where $\epsilon\ll 1$ is the weight associated with the noise term $\vecP_\text{\footnotesize noise}$. Moreover, for the purpose of illustrating the effectiveness of {the method} in identifying non-quantum-compatible data, we set $v> v_c$.  In our simulations, we set $\epsilon=0.01$ and $v=0.72>v_c$. However, as long as the given mixture lies outside $\AQ$ (and hence also outside $\Q$), the actual choices of $\epsilon\ll 1$ and $v\in(v_c,1]$ are  irrelevant. The only impact that these choices may have is the number of trials $\Ntot$ needed to falsify the hypothesis
\begin{displayquote}
 ``The observed data is compatible with a physical theory that is constrained to produce only the almost-quantum set of correlations."
\end{displayquote}
{\noindent}with the same level of confidence. Inspired by the experiments of Ref.~\cite{Christensen15}  where $\Ntot=10^5\sim 10^6$, we set in our simulations $\Ntot=10^6$. Note also that instead of $\AQ$, we can equally well choose another set of correlations that admits a semidefinite programming characterization, such as those described in Refs.~\cite{NPA,Moroder13}.

Since we are interested to model a nonlocal source that obeys the nonsignaling conditions of Equation~\eqref{Eq:NS},  there is no loss in generality by considering $\vecP_\text{\footnotesize noise}\in\NS$. To this end, let $\vecPExt$ be the $j$-th extreme point of the nonsignaling polytope~\cite{Barrett:PRA:2005}, then we may write $\vecP_\text{\footnotesize noise}=\sum_{j} p_j\vecPExt$ where $p_j$ is the weight associated with $\vecPExt$ in the convex decomposition of  $\vecP_\text{\footnotesize noise}$. We may thus write the nonlocal source of interest as:
\begin{equation}\label{Eq:vecPmixed}
	\vecPs:= (1-\epsilon)\vecP(v) + \epsilon\sum p_j\vecPExt.
\end{equation}
Finally, to simulate the raw data $\{(a_i,b_i,x_i,y_i)\}_{i=1}^{N}$ obtained in an $N$-trial Bell test for any given input distribution $P_{xy}$ and correlation $\vecP$, we make use of the MATLAB toolbox Lightspeed developed by Minka~\cite{lightspeed}.

\subsection{Simulations of Bell  Tests with an i.i.d. Nonlocal Source}
\label{Sec:iid}

\begin{figure}[h!tbp]
\begin{center}
\begin{tikzpicture}

\node (b) at (0,-1.5) {${ (a_i,b_i,x_i,y_i)}, i\in\{1,\ldots,\Nest,\Nest+1,\ldots,\Ntot\}$};
\node (c) at (-3,-3.9) {$\vecf$} ; 
\node (d) at (-1.3,-3.95) {\small{\textit{(Relative frequencies)}}} ; 
\node (e) at (-3,-5.9) {$\langle R(A,B,X,Y) \rangle \stackrel{\T}{\leq 1}$} ; 
\node (f) at (-0.4,-5.9) {\small{\textit{(Bell-like ineq.)}}} ; 
\node (g) at (3.2,-3.9) {$t = \prod_{i>\Nest} r_i$};
\node (h) at (3.2,-5.9) {$p\le \min \left\{\tfrac{1}{t},1\right\}$};

\draw[->] (b)--(c) node[midway, above, sloped] {\small{ Use ${ (a_i,b_i,x_i,y_i)},$ }};
\draw[] (b)--(c) node[midway, below, sloped] {\small{\qquad $ i\in\{1,\ldots,\Nest\}$}};
\draw[->] (c)--(e) node[midway,right] {\small{Minimize $\Dkl$}};
\draw[->,dashed] (b) -- (g) node[midway, above, sloped] {\phantom{abc}\small{Use ${ (a_i,b_i,x_i,y_i)},$}};
\draw[dashed] (b) -- (g) node[midway, below, sloped] {\small{$ {\Nest< i\le\Ntot}$\phantom{abc}}};
\draw[->,dashed] (f)--(g) node[midway, sloped] {};
\draw[->] (g)--(h) node[midway] {\small{$p$-value bound\phantom{a}}};
\end{tikzpicture}
\caption{\label{Fig:Simulation:Analysis} Flowchart summarizing the  steps involved in our application of the {prediction-based-ratio} method on the simulated data $\{(a_i,b_i,x_i,y_i)\}_{i=1}^{\Ntot}$ of a \emph{single} Bell test. In the first step, we separate the data into two sets, with the data collected from the first $\Nest$ trials serving as the training data while the rest is used for the actual hypothesis testing. Specifically, the training data is  used  to compute the relative frequencies $\vecf$ and to minimize the KL divergence $\Dkl$ with respect to the set of correlations  $\T\in\{\NS,\AQ\}$ associated, respectively, with the hypothesis of $\hNS$ and $\hAQ$. The correlation $\PKL{\T}\in\T$ that minimizes $\Dkl$ gives rise to a Bell-like inequality  with coefficients $\{R(A=a,B=b,X=x,Y=y)\}_{x,y,a,b}$. The remaining data is then used to compute $t=\prod_{i>\Nest} r_i$ where $r_i:=R(a_i,b_i,x_i,y_i)$. Finally, a $p$-value bound according to the hypothesis is obtained by computing $\min\{\frac{1}{t},1\}$.}
\end{center}
\end{figure}
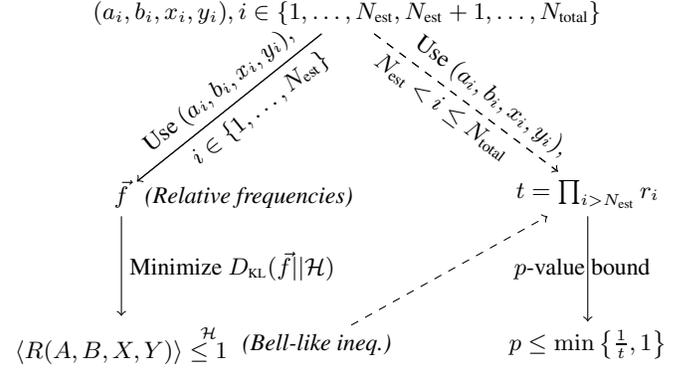

Let us begin with the case of i.i.d. trials, corresponding to a source of correlation that remains unchanged throughout the experiment, and where the inputs at each trial are independent of the inputs of the previous trials.  To this end, we first sample the weights $\{p_j\}_{j}$  uniformly from the interval $[0,1]$ and renormalize them such that $\sum_j p_j =1$. With our choice of $v=0.72$ and $\epsilon=0.01$, it is easy to find such a randomly generated correlation $\vecPs$ that lies outside $\AQ$. (Verifying that any given $\vecP$ is (not) in $\AQ$ can be carried out by solving a semidefinite program. Specifically, for any given correlation $\vecP$, if the maximal white-noise visibility $\nu$ such that $\nu\vecP + (1-\nu)\vecP_{\mathbb{I}}\in\AQ$ is smaller than 1, then $\vecP\not\in\AQ\supset\Q$, and hence outside $\Q$, otherwise $\vecP\in\AQ$.) For convenience, we denote by $\mathcal{P}$ the specific set of $\{p_j\}_j$ employed in our simulation of 500 Bell tests, each with $\Ntot=10^6$ trials. In Figure~\ref{Fig:Simulation:Analysis}, we summarize the steps involved in our analysis of the numerically simulated data using the {prediction-based-ratio} method. The resulting $p$-value upper bounds are summarized in Table~\ref{Table:iid}.

\begin{table}[h!tbp]
\caption{\label{Table:iid}  Summary of frequency distributions of the $p$-value upper bounds obtained from 500 numerically simulated Bell tests, each consists of $\Nest=10^6$ trials and assumes the same i.i.d. nonlocal source $\vecP(v,\epsilon,\{p_j\})$ of Equation~\eqref{Eq:vecPmixed} that lies \emph{outside} $\AQ$. The second and third row give, respectively, the frequency distributions according to the hypothesis associated with $\NS$ (nonsignaling) and $\AQ$ (almost-quantum). For these hypotheses,  the smallest $p$-value upper bound found among these 500 Bell tests are, respectively, 0.14 and 5.7$\times10^{-20}$. The second to the fifth column give, respectively, the fraction of simulated Bell tests having a $p$-value upper bound (for each hypothesis) that satisfies the given (increasing) threshold (e.g., $10^{-10}$ for the second column). Similarly, in the last column, we give the fraction of instances where the $p$-value upper bound obtained is trivial, i.e., exactly equals to 1.  The smaller the $p$-value upper bound, the less likely it is that a physical theory associated with the hypothesis produces the observed data.  Thus, the larger the value in the second (to the fourth) column, the less likely it is that the assumed physical theory holds true. In contrast, the larger the value in the rightmost column, the weaker the empirical evidence against the assumed theory is.}
\begin{center}
\begin{tabular}{c|cccc|c}\toprule
{\bf{$p$-Value Bound}} & {\bm{$\le 10^{-10}$}} & {\bm{$\le 10^{-4}$}} & {\bm{$\le 10^{-2}$}}& {\bm{$\le 10^{-1}$}} & {\bf Trivial} \\  \hline
$\NS$ & 0 & 0 & 0 & 0 &   97\% \\  
$\AQ$ &58\% & 85\% & 90\% & 93\% &   5.8\% \\ \hline
\end{tabular}

\end{center}
\end{table}

As expected, despite statistical fluctuations, the data does not suggest any obvious evidence against the nonsignaling hypothesis. In fact, among the 500 $p$-value bounds obtained, 97\% of them are trivial (i.e., equal to unity), while the smallest non-trivial $p$-value bound obtained is approximately 0.14. On the contrary, for the hypothesis test of the almost-quantum set of correlations, more than half of the simulated Bell tests give a $p$-value upper bound that is less than $10^{-10}$. Although there are also 5.8\% of these simulated Bell tests that give a trivial $p$-value bound according to the almost-quantum hypothesis, we see that the method generally works very well in falsifying this hypothesis. In fact, a separate calculation (not shown in the Table) shows that when we increase $\Ntot$ to $10^7$, all the 500 $p$-value upper bounds obtained according to the almost-quantum hypothesis are less than or equal to $10^{-10}$.

\subsection{Simulations of Bell tests with a non-i.i.d. Nonlocal Source}
\label{Sec:Niid}

In a real experiment, the assumption that the experimental trials are i.i.d is often far from justifiable, as that would require, for example, that the experimental setup remain as it is over the entire course of the experiment. As a result, we also consider here the case where the source that generates the data actually varies from one trial to another. To this end, for the $i$-th trial of the Bell test, we simulate according to the conditional outcome distributions:
\begin{equation}\label{Eq:vecPmixed:Niid}
	\vecP_i(v,\epsilon,n_i) = (1-\epsilon)\vecPv + \epsilon \vecP^\text{\tiny Ext}_{n_i},
\end{equation}
where $n_i=1,2,\ldots,24$ labels the \emph{single} nonsignaling extreme point used to mix with $\vecPv$ at this trial, cf. Equation~\eqref{Eq:vecPmixed}  with $p_j =1$ if $j=n_i$ but vanishes otherwise. Moreover, to facilitate a comparison with the i.i.d. case, before the $i$-th trial, we randomly pick $n_i$ according to the probability $P(n_i=j)=p_j$ where $p_j\in\mathcal{P}$ is exactly the probability employed in the simulation of Section~\ref{Sec:iid}. With this choice, the outcome distributions governed by the nonlocal source of Equation~\eqref{Eq:vecPmixed:Niid}  (for the $i$-th trial) averages to that of  Equation~\eqref{Eq:vecPmixed}  when the number of trials $\Ntot\to\infty$. Again, we follow the steps summarized in Figure~\ref{Fig:Simulation:Analysis}  to compute the relevant $p$-value upper bounds using the {prediction-based-ratio} method. The resulting $p$-value upper bounds are summarized in Table~\ref{Table:noniid}.

\begin{table}[h!tbp]
\caption{\label{Table:noniid}  Summary of frequency distributions of the $p$-value upper bounds obtained from 500 numerically simulated Bell tests. Each of these Bell tests involves $\Nest=10^6$ trials and each trial assumes a varying source $\vecP_i(v,\epsilon,n_i)$ of Equation~\eqref{Eq:vecPmixed:Niid}. For the hypothesis of $\hNS$ and $\hAQ$, associated with $\NS$ (second row)  and $\AQ$ (third row), respectively, the smallest $p$-value upper bound found among these 500 instances are 0.21 and 1.3$\times10^{-15}$. The significance of each column follows that described in the caption of Table~\ref{Table:iid}.
}
\begin{center}
\begin{tabular}{c|cccc|c}
\toprule
{\bf{$p$-Value Bound}} & {\bm{$\le 10^{-10}$}} & {\bm{$\le 10^{-4}$}} & {\bm{$\le 10^{-2}$}} & {\bm{$\le 10^{-1}$}} &   {\bf Trivial} \\ \hline
$\NS$ & 0 & 0 & 0 & 0 &   97\% \\ 
$\AQ$ & 17 & 59\% & 69\% & 72 &   24\% \\  \hline
\end{tabular}

\end{center}
\end{table}

As with the i.i.d. case, for these 500 simulated Bell tests, our application of the {prediction-based-ratio} method does not  lead to any obvious evidence against the nonsignaling hypothesis $\hNS$. However, for the hypothesis associated with the almost-quantum set $\AQ$, our results (last row of Table~\ref{Table:noniid})  give more than half of the $p$-value upper bounds that are less than $10^{-4}$ (accordingly, 17\% if we set the cutoff at $10^{-10}$). Although there are 24\% of these instances where the returned $p$-value upper bound for the same hypothesis is trivial,  we see that, as with the i.i.d. case, {the method} remains very effective in showing that the observed data cannot be entirely accounted for using a theory that is constrained to produce only almost-quantum correlations. In addition, as with the i.i.d. case, our separate calculation shows that the effectiveness of this method can be substantially improved when we increase $\Ntot$ to $10^7$: all the 500 $p$-value upper bounds obtained according to the almost-quantum hypothesis become less than or equal to $10^{-10}$.

\subsection{Application to Some Real Experimental Data}
\label{Sec:RealData}

Armed with the experience gained in the above analyses, let us now analyze the experimental results presented in Figure 3 of Ref.~\cite{Christensen15} using the {prediction-based-ratio} method. One of the goals of Ref.~\cite{Christensen15} was to experimentally approach the boundary of the quantum set of correlations in the two-dimensional subspace spanned by the two Bell parameters:
\begin{equation}
\begin{split}
	\SCHSH = E_{00} + E_{01} + E_{10} - E_{11}, \\
	\SCHSH' = -E_{00} + E_{01} + E_{10} + E_{1},
\end{split}	
\end{equation}
where $E_{xy}:=\sum_{a,b=0}^1 (-1)^{a+b}P(a,b|x,y)$ is the correlator. To this end, the Bell parameter $\SCHSH \cos{\theta} +   \SCHSH' \sin{\theta}$ for 180 uniformly-spaced values of $\theta\in\{\theta_1,\theta_2,\ldots,\theta_{180}\}\subset [0,2\pi)$ were estimated by performing the measurements presented in Appendix A of Ref.~\cite{Christensen15} on a two-qubit maximally entangled state.

Unfortunately, only the total counts for each combination of input-output $N_{a,b,x,y}$ (rather than the time sequences of raw data) given the value of $\theta$ are available~\cite{Brad:Private}. Therefore, in analogy with the analyses presented above, we use the relative frequencies obtained for $\theta_k$ as the training data to derive a {prediction-based ratio} (which corresponds to a Bell-like inequality)  for the hypothesis test using the data associated with $\theta_{k+1}$ (for the case of $k=180$, the hypothesis test uses the data associated with $\theta_1$). The analysis therefore essentially follows the steps outlined in Figure~\ref{Fig:Simulation:Analysis}, but with the computation of $t$ carried out using Equation~\eqref{Eq:T:simplified} instead, since we do not have the time sequences of raw data.
Moreover, to apply the {prediction-based-ratio} method, we {\em assume}, as with the numerical experiments reported earlier that the input distributions are uniform, i.e., $P_{xy}=\tfrac{1}{4}$ for all combinations of $x,y\in\{0,1\}$. A summary of the $p$-value upper bounds obtained from these 180 Bell tests is given in Table~\ref{Table:exp}.

For both hypotheses, approximately half of the $p$-value upper bounds obtained are trivial. At the same time, about the same fraction of the $p$-value bounds obtained are less than $10^{-2}$ (with the majority of them being less than $10^{-4}$). In fact, the smallest of the $p$-value upper bounds are remarkably small: 3.2$\times10^{-55}$ for the hypothesis of nonsignaling $\hNS$  and 2.7$\times10^{-55}$ for the hypothesis of almost-quantum $\hAQ$. These results strongly suggest that under the {\em assumption} that the measurement settings were \emph{randomly} chosen according to a uniform input distribution, it is extremely unlikely that a physical theory associated with each of these hypotheses can produce the observed relative frequencies. 

These conclusions that the observed data are incompatible with the fundamental principle of nonsignaling or with quantum theory (via the almost-quantum hypothesis), however, turn out to be {\em flawed}, as it was brought to our attention~\cite{Brad:Private} that during the course of the experiment, the measurement bases were not at all randomized---the measurements were carried out in blocks using the same combination of $(x,y)$ before moving to another. Why should this pose a problem? In the extreme scenario, if the measurement settings were fully correlated to some \emph{local} hidden variable, it is known that the the resulting correlation between measurement outcomes can violate the nonsignaling conditions of Equation~\eqref{Eq:NS},  see, e.g., Ref.~\cite{Putz2014}. 
Consequently, it is not surprising that in the {prediction-based-ratio} method (as well as any other methods employed for the statistical analysis of a Bell test),  the measurement inputs $(x_i,y_i)$ during the $i$-th trial, as discussed in Section~\ref{Sec:PBR}, ought to be randomly chosen. 

\begin{table}[h!tbp]
\caption{\label{Table:exp} Summary of frequency distributions of the $p$-value upper bounds obtained from the 180 Bell tests of Ref.~\cite{Christensen15} according to the hypothesis of $\hNS$ and $\hAQ$ (associated, respectively, with $\NS$, the second row, and $\AQ$, the third row) under the assumption that the measurement settings were randomly chosen according to a uniform distribution. The significance of each column follows that described in the caption of Table.~\ref{Table:iid}.
}
\begin{center}
\begin{tabular}{c|cccc|c}\toprule
{\bf{$p$-Value Bound}} & {\bm{$\le 10^{-10}$}} & {\bm{$\le 10^{-4}$}} & {\bm{$\le 10^{-2}$}} & {\bm{$\le 10^{-1}$}} &   {\bf{Trivial}} \\ \hline
$\NS$ & 38\% & 45\% & 48\% & 51\% &   48\% \\
$\AQ$ & 35\% & 44\% & 47\% & 49\% &   49\% \\ \hline
\end{tabular}

\end{center}
\end{table}

%%%%%
\section{Discussion}

As discussed in the last section, the conclusion that ``the experimental data of Ref.~\cite{Christensen15} show a violation of the nonsignaling principle" based on an erroneous application of the {prediction-based-ratio} method is unfounded. The results are nonetheless thought-provoking.  For example, suppose for now that we had access to the raw data for all trials. Since the analysis was flawed because of the nonrandomnization of measurement settings, one can imagine that---under the assumption that the trials are exchangeable---we first artificially randomize the hypothesis-testing trials to simulate the randomization of measurement settings in the experiment. Should we then expect to obtain $p$-value bounds with fundamentally different features? The answer is negative. The reason is that in our crude application of {the method}, only the number of counts $N'_{a,b,x,y}$ for each input-output combination matters, see Equation~\eqref{Eq:T:simplified}. In particular, the actual trials in which a particular combination of $(a,b,x,y)$ appears are irrelevant in such an analysis.

So, if one holds the view that the nonsignaling principle cannot be flawed, then one must come to the conclusion that ``should the measurement choices be randomized, it would be impossible to register the same number of counts $N'_{a,b,x,y}$ for each input-output combination''. A plausible cause for this is that the experimental setup suffered from some systematic drift during the course of the experiment, which is exactly a manifestation that the experimental trials are not i.i.d. It might then appear that a hypothesis test of the nonsignaling principle is hopeless in such a scenario. However, as mentioned above, the prediction-based-ratio method is applicable even for {\em non}-i.i.d. experimental trials. Indeed, as we illustrate in Section~\ref{Sec:Niid} (see, specifically Table~\ref{Table:noniid}), such fluctuations have not lead to any false positive in the sense of giving very small $p$-value upper bound according to the nonsignaling hypothesis.

More generally, as the above example of Section~\ref{Sec:RealData} illustrates, an unexpectedly small $p$-value upper bound according to the nonsignaling hypothesis may be a consequence that certain premises needed to perform a sensible Bell test are violated. In other words, an {\em apparent violation} as such does not necessarily pose a problem to any physical principle, such as the nonsignaling principle that is rooted in the theory of relativity. However, as nonlocal correlations also find applications in device-independent quantum information processing~\cite{Scarani12,Brunner14}, it is important to carry out such consistency checks alongside the violation of a Bell inequality before one  applies the estimated nonlocal correlation in any such protocols.

Of course, an unexpectedly small $p$-value upper bound according to the nonsignaling hypothesis could also be a consequence of mere statistical fluctuation. Indeed, our results in Section~\ref{Sec:iid} and Section~\ref{Sec:Niid} show that when a null hypothesis indeed holds {\em true}, it can still happen that one obtains a relatively small $p$-value upper bound (of the order of $10^{-1}$) even after a large number of trials ($\Ntot=10^6$). However, as explained in Appendix 1 of Ref.~\cite{Zhang2011},  if a null hypothesis is correct, the probability of obtaining a $p$-value upper bound smaller than $q$ with the {prediction-based-ratio} method is no larger than $q$. Indeed, in each of these instances, $p$-value upper bounds that are less than 10$^{-1}$ occur way less than 50 times among the 500 simulated experiments. In any case, this means that even though the prediction-based-ratio method already gets rids of the often unjustifiable i.i.d. assumption involved in such an analysis, the interpretation of the significance of a small $p$-value upper bound must still be carried out with care, as advised, for example, in Refs.~\cite{Nuzzo:Nature,Leek:Nature,pValue:ASA}.\newline

%%%%%
\section{Conclusion}

In this work, we revisited the prediction-based-ratio method developed~\cite{Zhang2011}---in the context of a Bell test---for performing hypothesis tests of LHV theories. We showed that with the two observations presented in Section~\ref{Sec:GeneralizedPBR}, the method can equally well be applied to perform hypothesis tests of \emph{other} physical theories, specifically those that are constrained to  produce correlations amenable to a semidefinite programming characterization. Prime examples of such theories include those that obey the principle of nonsignaling~\cite{Popescu1994}, those that satisfy the principle of macroscopic locality~\cite{Navascues:2010aa}, the so-called $v$-causal models~\cite{Bancal:2012aa}, as well as physical theories that are constrained to produce the almost-quantum set~\cite{Navascues:2015aa} or any other outer approximations~\cite{NPA,Moroder13,SLChen16} of the quantum set of correlations.

To illustrate the effectiveness of the method, we first numerically simulated 500 Bell tests using a hypothetical source of correlations that lies somewhat outside the almost-quantum set of correlations. We then applied {the method} to obtain a $p$-value upper bound according to both the almost-quantum hypothesis and the nonsignaling hypothesis for the simulated data obtained in each of these Bell tests. In the majority ($>90\%$) of these 500 instances, the $p$-value upper bound according to the almost-quantum hypothesis is less than $10^{-2}$. Since a $p$-value upper bound quantifies the evidence against the assumed (almost-quantum) theory given the observed data, these results show that in most of these simulated Bell tests, the data is unlikely to be explicable by the assumed theory. In a similar manner, we numerically simulated another 500 Bell tests using a hypothetical source that {\em varies} from one trial to another. Again, {the method} remained very effective (giving a $p$-value upper bound that is less than $10^{-2}$ for  69\% of the instances) in identifying the incompatibility between the observed data and the assumed (almost-quantum) theory in such a non-i.i.d. scenario. 

Finally, we applied the {prediction-based-ratio} method to the experimental data of Ref.~\cite{Christensen15}.  To this end, we assumed that the measurement settings were randomly chosen with uniform distributions. An application of {the method} under this assumption again led to very small $p$-value upper bounds ($10^{-4}$) for more than 40\% of the 180 Bell tests analyzed---not only for the almost-quantum hypothesis, but also for the nonsignaling hypothesis. Such a violation of the nonsignaling conditions, however, is apparent, as we learned after the analysis that the measurement settings were \emph{not} randomized during the course of the experiments, thereby invalidating one of the basic assumptions needed in the application of the {prediction-based-ratio} method. Nonetheless, as we remarked in the Discussion section, the analysis nevertheless unveils that the possibility of using the prediction-based-ratio method to identify a situation where a certain premise is needed to perform a proper Bell test, such as the randomization of settings, is invalidated.

{\em Note added}: While preparing this manuscript, we became aware of the work of Smania {\em et al.}~\cite{Smania:1801.05739}, which also discussed, among others, the implication of not randomizing the settings in a Bell test, and its relevance in quantitative applications.

\section*{Author Contributions}
Both authors contributed toward the computation of the numerical results and the preparation of the manuscript.

\section*{Funding}
This work is supported by the Ministry of Science and Technology, Taiwan (Grants No. 104-2112-M-006-021-MY3, 107-2112-M-006-005-MY2, 107-2627-E-006-001) and the National Center for Theoretical Science, Taiwan (R.O.C.).

\section*{acknowledgements}
Y.C.L. is grateful to Ad\'an Cabello, Bradley Christensen, Ehtibar Dzhafarov, Nicolas Gisin, Scott Glancy, Paul Kwiat, Jan-{\AA}ke Larsson, Denis Rosset, and Lev Vaidman for useful discussions.

\section*{Conflicts of Interest}
The authors declare no conflict of interest.

%%%%%%%%%%%%%%%%%%%%%%%%%%%%%%%%%%%%%%%%
%\bibliography{BellExpFiniteStatistics}

%

%%%%%%%%%%%%%%%%%%%%%%%%%%%%%%%%%%%%%%%%

\end{document}